\documentclass{article}
\usepackage{spconf,amsmath,graphicx}
\usepackage{bm, color}
\usepackage{amssymb}
\usepackage{algorithm}
\usepackage{comment}
\usepackage{algpseudocode}

\title{Diffusion model for gradient preconditioning in hyperspectral imaging inverse problems}
\name{Jonathan Monsalve$^{ *}$ and Kumar Vijay Mishra$^{\dagger}$\thanks{K. V. M. acknowledges support from the National Academies of Sciences, Engineering, and Medicine via the Army Research Laboratory Harry Diamond Distinguished Fellowship.}}
\address{$^{*}$Universidad de Investigación y Desarrollo, Bucaramanga 680002, Colombia \\
$^{\dagger}$United States DEVCOM Army Research Laboratory, Adelphi, MD 20783 USA
}

\begin{document}
%
\maketitle
\begin{abstract}
Recovering high-dimensional statistical structure from limited measurements is a
fundamental challenge in hyperspectral imaging, where capturing full-resolution data is often infeasible due to sensor, bandwidth, or acquisition constraints. A common workaround is to partition measurements and estimate local statistics—such as the covariance matrix—using only partial observations. However, this strategy introduces noise in the optimization gradients, especially when each partition contains few samples. In this work, we reinterpret this accumulation of gradient noise as a diffusion process, where successive partitions inject increasing uncertainty into the learning signal. Building on this insight, we propose a novel framework that leverages denoising diffusion models to learn a reverse process in gradient space. The model is trained to map noisy gradient estimates toward clean, well-conditioned updates, effectively preconditioning the optimization. Our approach bridges generative modeling and inverse problem solving, improving convergence and reconstruction quality under aggressive sampling regimes. We validate our method on hyperspectral recovery tasks, demonstrating significant gains in accuracy and stability over traditional optimization pipelines.
\end{abstract}
\begin{keywords}
Diffusion models, Gradient preconditioning, hyperspectral imaging, covariance estimation
\end{keywords}
\section{Introduction}
\label{sec:intro}
Hyperspectral imaging (HSI) captures detailed spectral information across numerous wavelength bands, providing rich 3D data structures commonly referred to as hyperspectral cubes. While this level of spectral resolution enables precise material identification and analysis, acquiring full-resolution hyperspectral data is often costly, time-consuming, and limited by sensor constraints. As a result, many practical systems rely on collecting random low-dimensional projections of the scene rather than capturing the full hyperspectral cube directly\cite{MioAp}.

In many HSI applications, it is more crucial to estimate statistical structures such as the spectral covariance matrix than to reconstruct the full image. The covariance matrix encodes critical information about spectral variability, enabling tasks like anomaly detection, classification, and dimensionality reduction. Estimating the covariance matrix from random linear projections has thus emerged as an effective alternative to full-data reconstruction, often formulated as a convex optimization problem\cite{Besson2008, invariance2}.

To mitigate the ill-conditioning inherent in this inverse problem, a common strategy is to divide the data into partitions and perform multiple independent projections. This approach enhances the identifiability of the covariance structure by introducing diverse sensing views, all while keeping the total number of measurements fixed. However, it also introduces noise due to the reduced number of samples per partition. This trade-off leads to a degradation in the optimization gradient, which accumulates noise in a structured manner reminiscent of a forward diffusion process\cite{MioAp, MioCov, MioCov2}.

Diffusion models, originally developed in generative modeling, are based on the principle of gradually corrupting data via a forward stochastic process and then learning to reverse this corruption through a neural network-based denoising procedure\cite{Diffusion2,ho2020denoising}. Motivated by this connection, we propose a novel framework that uses diffusion models to precondition the gradient of the covariance estimation problem. Specifically, we treat the noise induced by partitioning as part of a diffusion-like process and apply learned denoising steps to restore clean gradient directions during optimization. This approach enhances convergence and improves the fidelity of the recovered covariance structure.

Extensive simulations demonstrate that the proposed gradient preconditioning approach significantly enhances recovery performance. In particular, applying diffusion-based denoising to the gradient leads to improvements of up to 50\% reduction in MSE(Mean Squared Error) compared to the baseline. This substantial reduction in error highlights the effectiveness of noise-aware gradient correction during optimization. Furthermore, an eigenstructure analysis of the recovered covariance matrices reveals that our method accurately reconstructs up to two additional leading eigenvectors compared to baseline techniques such as Gaussian-filtered gradient descent. This improved spectral fidelity is especially valuable in hyperspectral imaging tasks, where accurate subspace recovery is critical for downstream applications like material identification and anomaly detection.

\section{System Model}
\label{sec:system}
Let $\mathbf{X} = [\mathbf{x}_1, \ldots, \mathbf{x}_n] \in \mathbb{R}^{l \times n}$ be a data matrix, where each column $\mathbf{x}_j \in \mathbb{R}^l$ for $j = 1, 2, \ldots, n$ represents an independent realization of a zero-mean Gaussian random vector with covariance matrix $\bm{\Sigma}$. That is, conditioned on $\bm{\Sigma}$, the distribution of $\mathbf{x}$ is given by $
    f(\mathbf{x}|\bm{\Sigma}) = \pi^{-l/2} |\bm{\Sigma}|^{-l/2} \text{etr}\left(-\frac{1}{2}\bm{\Sigma}^{-1} \mathbf{x}\mathbf{x}^T\right)$
where $\text{etr}(\cdot)$ denotes the exponential of the trace. Under this model, the maximum likelihood estimator (MLE) of the covariance matrix is the sample covariance matrix
    $\mathbf{S} = \frac{1}{n}\mathbf{X}\mathbf{X}^T = \frac{1}{n} \sum_{j=1}^{n} \mathbf{x}_j \mathbf{x}_j^T,$
with $\bm{\Sigma} = \mathbb{E}[\mathbf{S}]$ and $\bm{\Sigma} \in S_{++}^{l \times l}$, the set of positive definite matrices of size $l \times l$. In many practical scenarios, we do not observe the full high-dimensional signal $\mathbf{x}_j$, but rather a compressed version obtained via a linear projection:
\begin{equation}
    \mathbf{Y} = \mathbf{P}^T \mathbf{X} + \mathbf{N} = [\mathbf{P}^T \mathbf{x}_1, \ldots, \mathbf{P}^T \mathbf{x}_n] + \mathbf{N},
    \label{eq:randomnoise}
\end{equation}
where $\mathbf{Y} \in \mathbb{R}^{m \times n}$ contains the projected measurements $\mathbf{y}_j \in \mathbb{R}^m$, $\mathbf{P} \in \mathbb{R}^{l \times m}$ is a fixed projection matrix with $m < l$, and $\mathbf{N} \in \mathbb{R}^{m \times n}$ is an additive noise matrix with entries $N_{i,j} \sim \mathcal{N}(0, \sigma_N^2)$ i.i.d. The sample covariance matrix of the compressed observations is given by:
\begin{equation}
    \mathbf{\tilde{S}} = \frac{1}{n} \mathbf{Y} \mathbf{Y}^T = \frac{1}{n} (\mathbf{P}^T \mathbf{X} + \mathbf{N})(\mathbf{P}^T \mathbf{X} + \mathbf{N})^T,
    \label{eq:5}
\end{equation}
where $\mathbf{\tilde{S}} \in \mathbb{R}^{m \times m}$. Since $\mathbf{x} \sim \mathcal{N}(0, \bm{\Sigma})$ and $\mathbf{P}$ is fixed, the projected vectors $\{\mathbf{y}_j\}_{j=1}^{n}$ are also Gaussian with zero mean and covariance $\mathbf{P}^T \bm{\Sigma} \mathbf{P} + \sigma_N^2 \mathbf{I}$, i.e.,$
    \mathbf{y} \sim \mathcal{N}(0, \mathbf{P}^T \bm{\Sigma} \mathbf{P} + \sigma_N^2 \mathbf{I}).$
    
A common strategy to estimate $\bm{\Sigma}$ from $\mathbf{\tilde{S}}$ is to minimize the Frobenius norm of the difference between the empirical covariance of the projected data and the projection of the covariance estimate\cite{ Besson2008, invariance2, Romero2016Compressive}:
\begin{equation}
    \bm{\Sigma}^* = \underset{\bm{\Sigma} \in \mathcal{D}}{\arg\min} \; \| \mathbf{\tilde{S}} - \mathbf{P}^T \bm{\Sigma} \mathbf{P} \|_F^2 + \tau \psi(\bm{\Sigma}),
    \label{eq:opt1}
\end{equation}
where $\psi(\cdot)$ is a convex regularization function (e.g., promoting low-rank or Toeplitz structure), $\tau$ is a regularization parameter, and $\mathcal{D}$ is a convex closed set (e.g., positive semi-definite matrices). In order to ease this problem, Monsalve et al.\cite{MioCov} proposed splitting the signal to have multiple projections of the covariance matrix. Let us randomly partition the data set $\mathbf{X} \in \mathbb{R}^{l \times n}$ into $p$ disjoint subsets $\{\mathbf{X}_i\}_{i=1}^p$, each containing $b = n/p$ randomly selected columns of $\mathbf{X}$. Specifically, let $\Omega$ be a random permutation of $\{1, 2, \ldots, n\}$. Then, define the index set for the $i$-th subset as: $\Omega_i = \{\Omega_{(i-1)b + 1}, \ldots, \Omega_{ib}\},$
and construct the corresponding data block as: $\mathbf{X}_i = [\mathbf{x}_{j}]_{j \in \Omega_i}.$ 

Each $\mathbf{X}_i \in \mathbb{R}^{l \times b}$ is then projected into a lower-dimensional space using a distinct random projection matrix $\mathbf{P}_i \in \mathbb{R}^{l \times m}$, producing compressive measurements:
\begin{equation}
    \mathbf{Y}_i = \mathbf{P}_i^T \mathbf{X}_i + \mathbf{N}_i,
    \label{eq:partition}
\end{equation}
where $\mathbf{N}_i$ models additive Gaussian noise. The main idea behind this partition approach is that the covariance matrix of the process does not change among the partitions, even though the actual realization does. Hence, we have multiple measurements $\mathbf{\tilde{S}}_i = \frac{1}{b} \mathbf{Y}_i\mathbf{Y}_i^T \approx \frac{1}{n}\mathbf{P}\bm{\Sigma}\mathbf{P}^T$ instead of a single low-dimensional projection as in \eqref{eq:opt1}. Note that this approach does not increase the number of measurements as it captures low-dimensional projections of these disjoint subsets. Hence, the resulting optimization problem to recover the covariance matrix from multiple partitions is given by
\begin{equation}
	\begin{aligned}
	\bm{\Sigma}^* = &\underset{\bm{\Sigma} \in \mathcal{D}}{\text{ argmin}}
	& & \sum_{i=1}^{p} ||\mathbf{\tilde{S}}_i-\mathbf{P}_i^T\bm{\Sigma}\mathbf{P}_i||_F^2 + \tau \psi (\bm{\Sigma}).
	\end{aligned}
	\label{eq:opt3}
\end{equation}

This partitioning strategy improves the invertibility of the Fisher Information Matrix \cite{MioCov}, leading to better estimation performance compared to using a single global projection. However, it also introduces a variance term in the gradient, given by
\begin{equation}
\text{Error}[\nabla \tilde{f}(\bm{\Sigma})] = -\sum_{i=1}^p \mathbf{P}_i \mathbf{P}_i^T \mathbf{R}_i \mathbf{P}_i \mathbf{P}_i^T,
\label{eq:error}
\end{equation}
where $\nabla \tilde{f}(\bm{\Sigma})$ denotes the gradient of the objective function in \eqref{eq:opt3}. As the number of partitions $p$ increases, the problem becomes better conditioned due to the diversity of measurements, but at the cost of amplifying the gradient error, which accumulates proportionally to $p$.

\section{Gradient preconditioning via Diffusion models}
Diffusion models are a class of generative models that learn to sample from complex data distributions by reversing a parameterized noising process. They operate in two stages: a \textit{forward diffusion process}, which gradually adds Gaussian noise to the data, and a \textit{reverse denoising process}, learned via a neural network, that reconstructs the data from noise\cite{ho2020denoising, Diffusion2} or from noisy versions of the image\cite{Liu_2024_CVPR}. 
To formalize this analogy, consider the sequence of gradient estimates:
\begin{equation}
\nabla \tilde{f}^{(1)}(\bm{\Sigma}) \rightarrow \nabla \tilde{f}^{(2)}(\bm{\Sigma}) \rightarrow \cdots \rightarrow \nabla \tilde{f}^{(p)}(\bm{\Sigma}),
\end{equation}
where \(\nabla \tilde{f}^{(k)}\) denotes the gradient of the objective function computed using \(k\) partitions, which introduces noise due to sample reduction. For notational simplicity, we will write \(\nabla \tilde{f}^{(k)}\) instead of the full expression \(\nabla \tilde{f}^{(k)}(\bm{\Sigma})\) throughout the remainder of the text. Each step adds a noise increment \(\Delta \mathbf{G}_k = -\mathbf{P}_k \mathbf{P}_k^T \mathbf{R}_k \mathbf{P}_k \mathbf{P}_k^T\), with growing variance as partition size shrinks. In the forward process, a data point $\nabla \tilde{f}^{(k)} \sim p_{\text{data}}(\nabla \tilde{f})$ is transformed into a sequence $\{\nabla \tilde{f}^{(k)}\}_{k=1}^p$ through a Markov chain defined as:
\begin{equation}
q(\nabla \tilde{f}^{(k)} \mid \nabla \tilde{f}^{(k-1)}) = \mathcal{N}(\nabla \tilde{f}^{(k)}; \sqrt{1 - \beta_k} \, \nabla \tilde{f}^{(k-1)}, \beta_t \mathbf{I}),
\label{eq:markovchain}
\end{equation}
where $\beta_k \in (0, 1)$ is a variance schedule controlling the noise magnitude at each timestep $k$. Additionally lets two additional terms to simplify the formulation of the diffusion process. The first is the signal retention factor at each step:
$\alpha_k = 1 - \beta_k,$
and the second is the cumulative product of these retention factors up to step $k$: $ 
\bar{\alpha}_k = \prod_{i=1}^k \alpha_i.$ As stated in \eqref{eq:error} the covariance matrix is disturbed by an error that increases as $t\xrightarrow{}p$. It was shown in \cite{MioCov} that the error term follows a Gaussian distribution. The reverse process seeks to learn the backward transition probabilities $p_\theta(\nabla \tilde{f}^{(k-1)} \mid \nabla \tilde{f}^{(k)})$ to reverse the diffusion and generate noiseless samples. These transitions are modeled as Gaussian:
\begin{equation}
\begin{aligned}
&p_\theta(\nabla \tilde{f}^{(k-1)} \mid \nabla \tilde{f}^{(k)}) = \\&\mathcal{N}(\nabla \tilde{f}^{(k-1)}; \boldsymbol{\mu}_\theta(\nabla \tilde{f}^{(k)}, k), \mathbf{C}_\theta(\nabla \tilde{f}^{(k)}, k)),
\end{aligned}
\end{equation}
where $\boldsymbol{\mu}_\theta$ and $\mathbf{C}_\theta$ are parameterized by a neural network. Training minimizes a variational bound on the negative log-likelihood or an equivalent denoising score-matching objective. We employ a U-Net-based convolutional neural network to learn the reverse denoising process, conditioned on the diffusion step $p$. The scalar $p$ is encoded using sinusoidal positional embeddings and injected into the encoder features via a small multilayer perceptron\cite{positionalsembeddings}. The network follows a standard U-Net structure with two downsampling and upsampling stages, including skip connections. To preserve the symmetric structure of the gradient matrices, a final symmetrization layer computes $0.5(\mathbf{X}+\mathbf{X}^T)$ at the output. This step is necessary because the gradient of the objective function, like the covariance matrix itself, is symmetric due to its derivation from a positive semi-definite structure. The network is trained to minimize the $\ell_2$ norm between the estimated gradient noise and the true error defined in equation \eqref{eq:error}. The training loss is given by:
\begin{equation}
L_{\text{simple}} = \mathbb{E}_{k, \text{Error}[\nabla \tilde{f}], \nabla \tilde{f}^{(0)}} \left\| \text{Error}[\nabla \tilde{f}^{(k)}] - \epsilon_{\theta}(\nabla \tilde{f}^{(k)}, k) \right\|^2,
\end{equation}
where $\epsilon_{\theta}(\nabla \tilde{f}^{(k)}, k)$ denotes the network estimate of error at noise level $k$.

\begin{figure}
    \centering
    \includegraphics[width=0.99\linewidth]{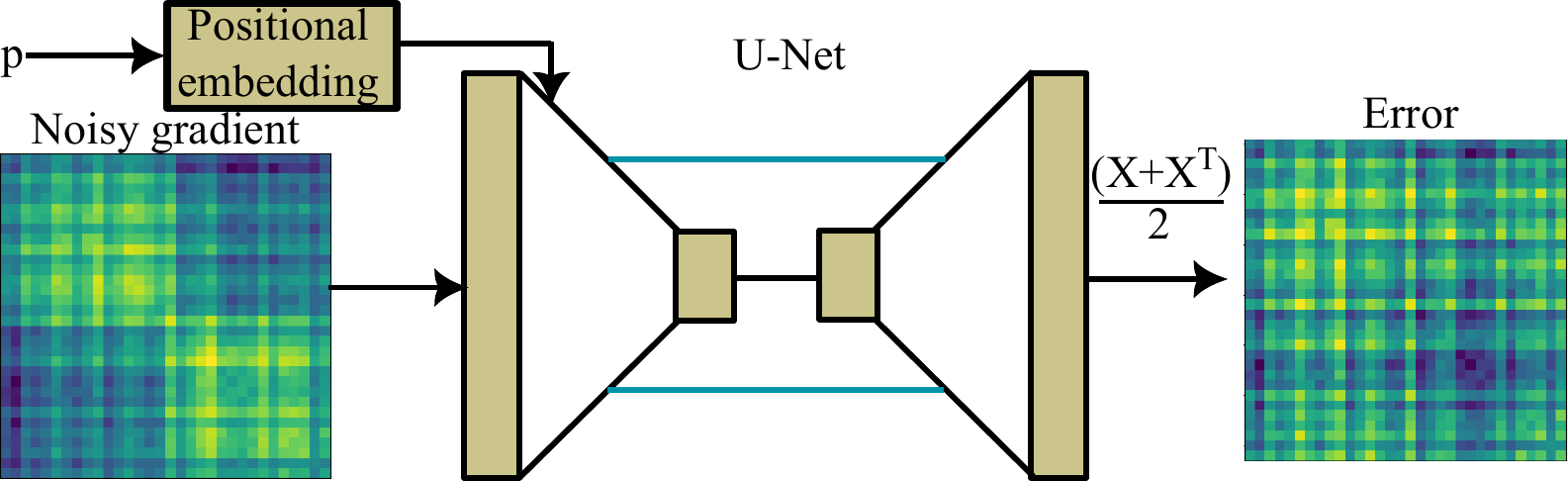}
    \caption{Schematic overview of the denoising architecture: The noisy gradient is processed through a U-Net, followed by a symmetrization layer that ensures the output is symmetric, approximating the structured noise component to be removed}
    \label{fig:enter-label}
\end{figure}
\begin{algorithm}[t]
\caption{Gradient Preconditioning via Diffusion model}
\begin{algorithmic}[1]
\State \textbf{Input:}  $\bm{\Sigma}^0 \in \mathcal{D}$,  $\{\alpha_k\}_{k=1}^p$, $\lambda_0$, $\tau$
\State \textbf{Initialize:} $j \gets 0$
\While{stopping criterion not satisfied}
    \State Choose step size $\lambda_j > 0$ \Comment{e.g., Armijo line search}
    \State Compute noisy gradient: $\nabla \tilde{f}^{(p)} \gets \nabla f(\bm{\Sigma}^j, \tau)$
    \For{$k = p, \dots, 1$} //  \textbf{sampling}
      \State $\mathbf{z} \sim \mathcal{N}(\mathbf{0}, \mathbf{I})$ if $k > 1$, else $\mathbf{z} = \mathbf{0}$
      \State $\nabla \tilde{f}^{k-1} \gets \frac{1}{\sqrt{\alpha_k}} \left( \nabla \tilde{f}^k - \frac{1 - \alpha_k}{\sqrt{1 - \bar{\alpha}_k}} \, \epsilon_\theta(\nabla \tilde{f}_k, k) \right) + \sigma_k \mathbf{z}$
    \EndFor
    
    \State Update estimate: $\bm{\Sigma}^{j+1} \gets P_{\mathcal{D}}(\bm{\Sigma}^j - \lambda_j  \nabla \tilde{f}^0)$
    \State $j \gets j + 1$
\EndWhile
\end{algorithmic}
\label{alg:mio}
\end{algorithm}
\section{Numerical results}
We evaluate our method using the OHID-1 dataset, a hyperspectral image collection built from Orbita satellite data over Zhuhai City, China\cite{Deng2024}. Each image in OHID-1 contains 32 spectral bands and has a spatial resolution of \(512 \times 512\) pixels. The dataset includes 10 images; for our experiments, we used 9 images for training and reserved 1 image for testing. 

\begin{figure}
    \centering
    \includegraphics[width=0.99\linewidth]{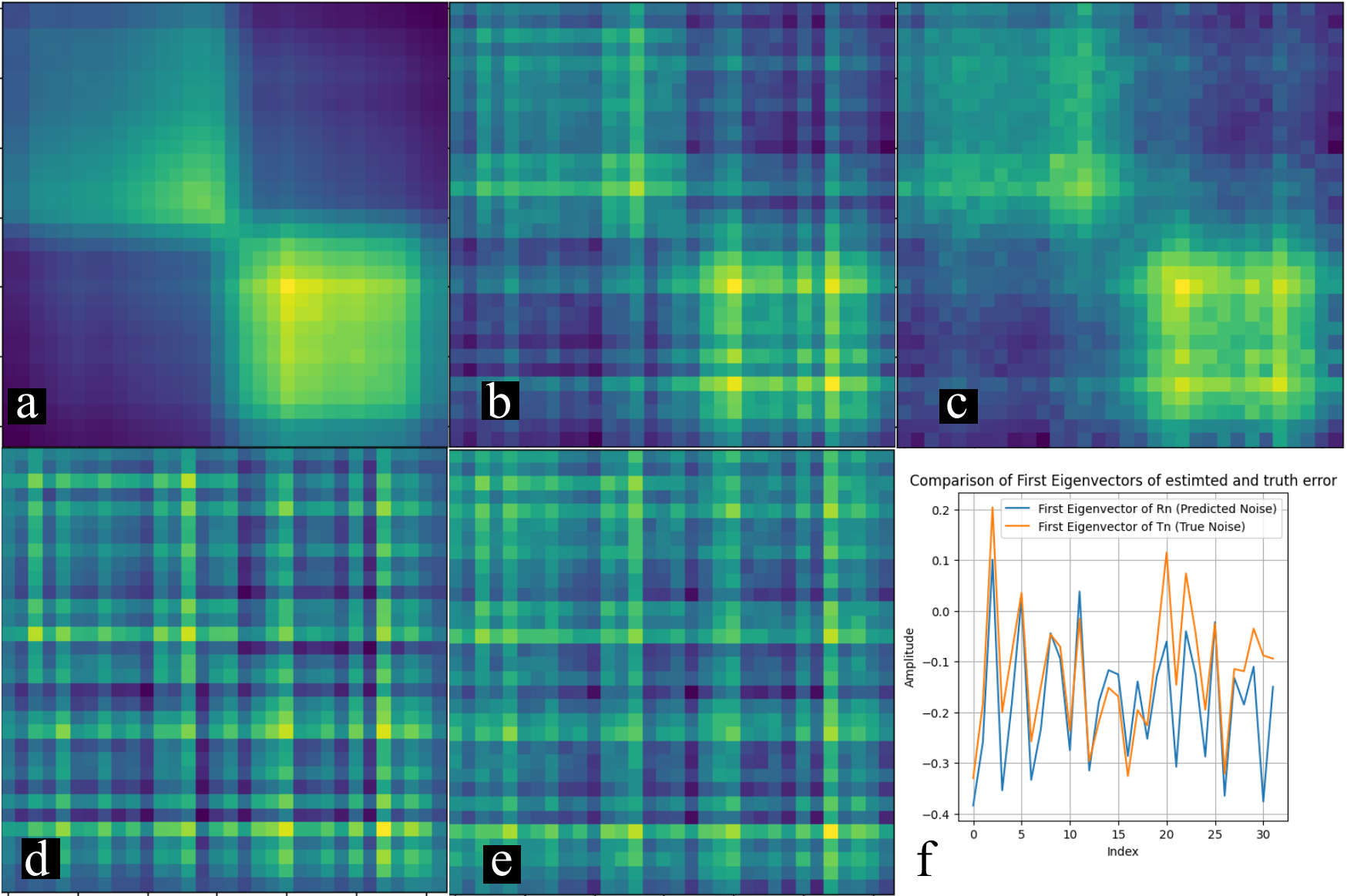}
    \caption{Visual analysis of the diffusion-based gradient denoising. (a) Clean gradient computed with a single partition ($p = 1$). (b) Noisy gradient with $p = 256$ partitions, exhibiting structured noise. (c) Denoised gradient estimated by the diffusion model. (d) Ground truth noise error. (e) Predicted noise error. (f) Comparison of the first eigenvectors of the true and predicted error, showing strong alignment.}
    \label{fig:noise}
\end{figure}

To generate training samples for the diffusion model, we set a maximum number of partitions \(p_{\max} = 1024\). For each training instance, a random number of partitions \(p \leq p_{\max}\) was selected, and the corresponding gradient noise term was computed according to the error formulation in \eqref{eq:error}. This process emulates the forward diffusion process, where increasing \(k\) introduces structured noise into the gradient due to reduced sample size per partition. The noisy gradient \(\nabla \tilde{f}^{(k)}\) serves as the input to the denoising network. The U-Net was trained to estimate the noise component added to the clean gradient, effectively learning the reverse denoising transitions of the diffusion process. Figure~\ref{fig:noise} illustrates the performance of the proposed diffusion-based denoising model on a single instance of a noisy gradient. Subfigure (a) displays the clean gradient matrix, obtained using a single partition ($p = 1$). Subfigure (b) shows a noisy gradient realization corresponding to $p = 256$, where the noise introduces structured artifacts due to the symmetry constraint. Subfigure (c) presents the denoised gradient produced by the diffusion model after the reverse sampling process. Although some randomness remains, the structured noise patterns are significantly suppressed. Subfigures (d) and (e) visualize the ground truth error and the predicted error, respectively. Finally, subfigure (f) compares the first eigenvectors of the predicted and true error matrices, showing strong alignment and suggesting the denoiser captures the principal direction of the error accurately.

In this experiment, we evaluate the effectiveness of diffusion-based gradient denoising for covariance matrix estimation algorithm \ref{alg:mio}. We set the number of partitions to $p=256$, which corresponds to the highest noise level in our setup. For each partition, random projection matrices of size $32\times9$ were used, resulting in approximately compression 70\% relative to the spectral dimension.

\begin{figure}
    \centering
    \includegraphics[width=0.99\linewidth]{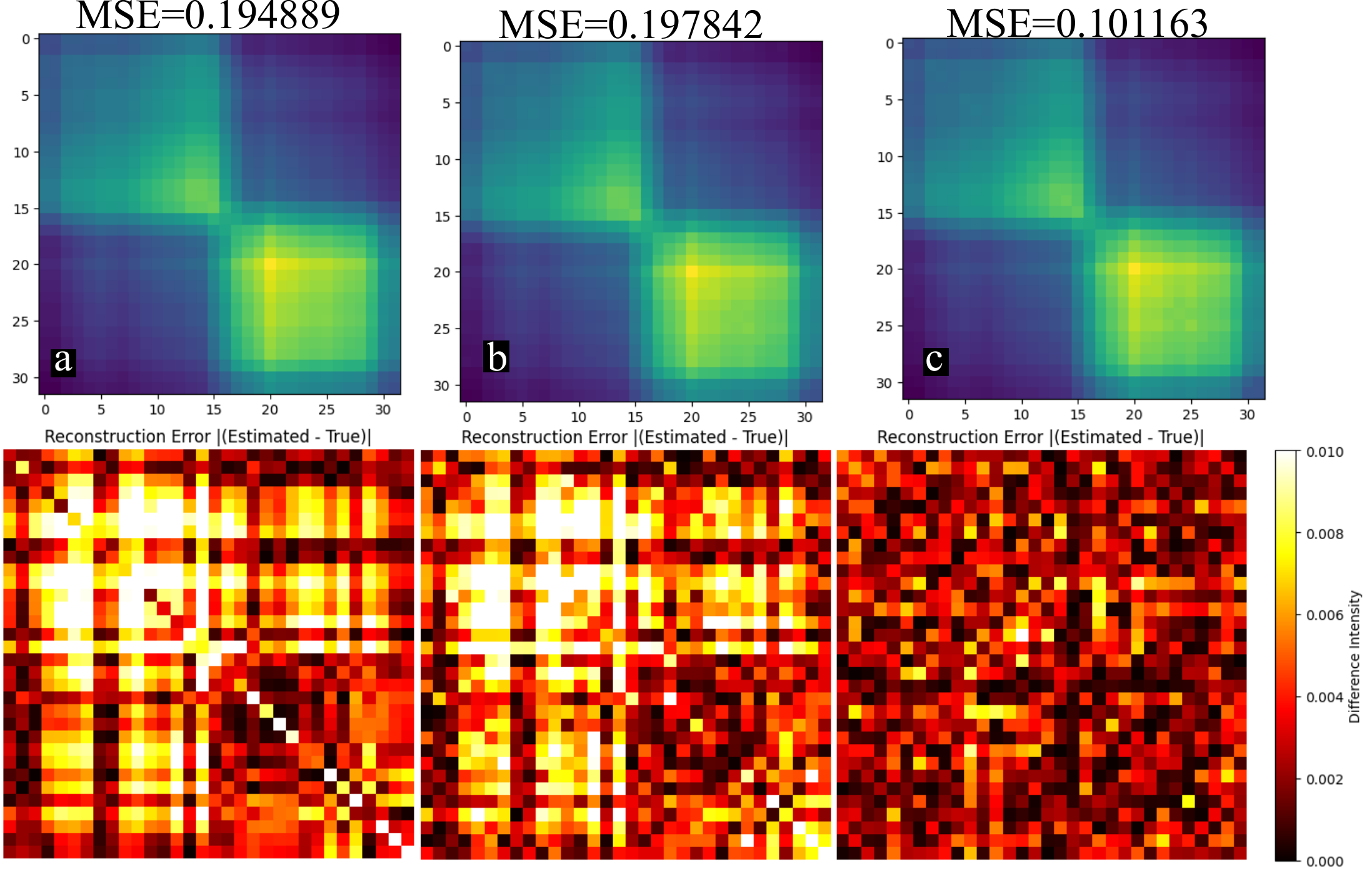}
    \caption{Comparison of covariance reconstruction using different preconditioning strategies in the gradient descent algorithm. (a) Gaussian-filtered gradient, (b) raw noisy gradient (no filtering), and (c) the proposed diffusion-based preconditioning. Top row: reconstructed covariance matrices. Bottom row: absolute reconstruction error $|\bm{\Sigma} - \bm{\Sigma}^*|$, normalized using a fixed intensity scale. Mean Squared Error (MSE) is reported above each case. The diffusion-based preconditioning yields the lowest reconstruction error, both visually and quantitatively.}
    \label{fig:error}
\end{figure}

As the optimization backbone, we adopted the projected gradient descent algorithm introduced in \cite{MioCov}, which ensures that the iterates remain within the feasible set of symmetric positive semi-definite (PSD) matrices. In their original formulation, the authors proposed a simple Gaussian filtering scheme to suppress gradient noise. We compare our approach directly against this Gaussian filtering baseline as well as the unfiltered case.
To evaluate the impact of different preconditioning strategies on the quality of covariance reconstruction, we compare three approaches within the gradient descent framework. Figure~\ref{fig:error} illustrates the estimated covariance matrices (top row) and their corresponding absolute reconstruction errors (bottom row) under three different setups: a) A Gaussian filter applied to the noisy gradient prior to the update step. b) No preconditioning applied, i.e., directly using the raw noisy gradient, c) and our proposed diffusion-based preconditioning, where the noisy gradient is denoised using the trained diffusion model.

The visual comparison highlights that the proposed method achieves a significantly cleaner reconstruction, particularly in structured regions of the matrix. This is further confirmed by the Mean Squared Error (MSE) values reported above each matrix. While the Gaussian filter provides a slight improvement over the raw gradient, it fails to preserve finer structural details. In contrast, the diffusion-based method achieves a more accurate and structured denoising, leading to the best overall reconstruction.

\section{Conclusions}
In this work, we proposed a diffusion-based gradient preconditioning strategy to enhance covariance matrix estimation under extreme compression settings. By modeling the structured noise in the stochastic gradient as a denoising problem, we trained a U-Net to learn and predict the noise component over a diffusion process, significantly improving the quality of the estimated gradients.

We validated our approach on the OHID-1 hyperspectral dataset, using 90\% of the images for training and 10\% for evaluation. Compared to standard approaches including raw gradient descent and Gaussian filtering We observed a consistent improvement in the quality of the reconstructed covariance matrices. Notably, our method reduced the mean squared error by nearly 50\% relative to unfiltered estimates and produced reconstructions with fewer structural artifacts.

The proposed framework is flexible and can be extended to other structured estimation problems where gradient noise is non-trivial and spatially correlated. Future work will explore extensions to online settings and adaptive noise scheduling to further enhance convergence and generalization.
\bibliographystyle{IEEEbib}
\bibliography{refs}

\end{document}